\documentclass[journal]{IEEEtran}
\usepackage{cite}

\ifCLASSINFOpdf
   \usepackage[pdftex]{graphicx}
   \DeclareGraphicsExtensions{.pdf,.jpeg,.png}
\else
\fi
\usepackage{amsmath}
\begin{document}
%
\title{Potential of non-conventional superconductors for particle accelerator cavities}
%
%
%

\author{Akira~Miyazaki
\thanks{A. Miyazaki is with CNRS/IN2P3/IJCLab Universit\'{e} Paris-Saclay}
}

%
%

\markboth{submitted to IEEE TRANSACTIONS ON APPLIED SUPERCONDUCTIVITY, NOV 2023}%
{A.~Miyazaki: Potential of non-conventional superconductors for particle accelerator cavities}
%



\maketitle

\begin{abstract}
Superconducting RadioFrequency (SRF) cavities have been developed for modern particle accelerator projects in the world.
These cavities are mainly made of bulk niobium operated in superfluid or normal fluid helium.
Due to the price increase in niobium material as well as helium,
superconductors beyond the niobium are investigated for more sustainable accelerators.
In this paper, we consider the most fundamental loss mechanism of microwaves inside various superconductors caused by thermally excited quasi-particles in the Meissner state.
\end{abstract}


%
\IEEEpeerreviewmaketitle

\section{Introduction}
\IEEEPARstart{S}{uperconducting} RadioFrequency (SRF) cavities have been developed for the particle accelerator application.
The SRF technology is versatile and can be applied to different accelerators, such as energy-frontier circular colliders~\cite{VenturiniDelsolaro:2856428}, high-intensity electron accelerators~\cite{Angal-Kalinin_2018}, and linear accelerators for protons and heavy ions~\cite{osti_1346823}.
Construction of the SRF accelerators has been realized mainly with bulk niobium as cavity material.
Such conventional cavities made of niobium are operated inside superfluid or normal fluid helium depending on its operating frequency.
The increasing costs of the material and cryogenic system including helium gas gave rise to the fundamental question of the sustainability of the present SRF technology.
One of the research directions is to explore new superconducting materials potentially operated at higher temperatures.

 

The quest for new superconductors available for accelerator cavities has been widely performed in the SRF community.
One classic example is thin film niobium instead of bulk niobium in order to reduce the material cost~\cite{niobium:2319900}.
By depositing niobium film on copper substrates, thermal stability is also dramatically improved.
Another established technology is Nb$_{3}$Sn coating on bulk niobium substrates~\cite{Posen_2017}.
This improves the surface resistance such that the performance at 4~K becomes equivalent to that of niobium at 2~K;
therefore, one can dramatically reduce the cryogenic cost.
This technology opened the possibility of using cryo-coolers instead of liquid helium systems~\cite {PhysRevAccelBeams.26.044701}.

The previous studies have been based on an approximate formula on the surface resistance of the conventional superconductors caused by thermally excited quasi-particles, given by~\cite{halbritter70}
\begin{equation}
R_{\rm BCS} = \frac{A \omega^{2}}{T} \exp{\left(-\frac{\Delta}{k_{\rm B}T} \right)}, \label{eq:BCS}
\end{equation}
where $A$ and $\Delta$ are the material constants often determined by data fitting.
This is called the BCS resistance in the SRF community.
For example, $\Delta$ of Nb$_3$Sn is around twice larger than that of niobium and therefore the performance at higher temperatures can be comparable to niobium at lower temperatures.
This formula is the starting point of almost all the arguments in the SRF community.

Equation~(\ref{eq:BCS}) was explicitly derived for the simplest one-gap s-wave superconductors, such as niobium and Nb$_3$Sn.
For more non-conventional superconductors, such as multi-gap and d-wave superconductors, 
one needs to consider a more fundamental theory to compare different materials.
The aim of this paper is to review such a theory and to obtain a more unified view of the surface resistance of general superconducting materials.

Note that $R_{\rm BCS}$ is only one of the causes of the surface resistance.
Heuristically, the so-called residual resistance $R_{\rm res}$ is added to $R_{\rm BCS}$ to make total surface resistance.
Part of this term can be explained by adding sub-gap states with a phenomenological Dynes parameter~\cite{Gurevich17}, which is related to some magnetic impurities in the material~\cite{PhysRevB.94.144508}.
For simplicity, in this work, we do not consider various sources of $R_{\rm res}$.
We briefly address this point in section~\ref{sec:residual}.


\section{Theory}
\begin{figure}[!t]
\centering
\includegraphics[width=1.8in]{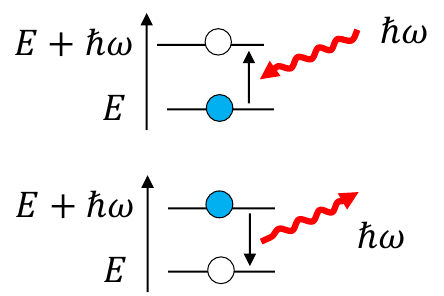}
\caption{Schematic of RF absorption and emission by quasi-particles. The quantum states of the quasi-particles are quantized but the RF fields are classical in this formulation (semi-classical theory)}
\label{fig:RF_absorption}
\end{figure}

The BCS resistance Eq.~(\ref{eq:BCS}) can be generalized when we go one step back in its derivation in the linear response theory~\cite{mattis58, abrikosov59, PhysRev.156.470}.
The RF loss mechanism in the Meissner state can be modelled by the net RF absorption by quasi-particles~\cite{halbritter74} as shown in Fig.~\ref{fig:RF_absorption}.
Some quasi-particles absorb RF photons of angular frequency $\omega$ while the other emits them.
Integrating over all the quasi-particles leads to the optical response against the applied RF fields.
The quantum states of the quasi-particles can be given by a model of the superconductor described by the Green function method~\cite{kopnin01}.

If we disregard non-locality, the surface resistance is given by
\begin{equation}\label{eq:Rs}
R_{\rm s}(T) = \frac{\mu_0\omega^2\lambda^3}{2} \sigma_1 (T),
\end{equation}
where $\lambda$ is the penetration depth and $\sigma_1(T)$ is the real part of the optical conductivity
\begin{equation}\label{eq:sigma1}
   \sigma_1(T)=\frac{\sigma_{\rm n}}{\hbar\omega} \left(1-e^{-\omega/T} \right) \int_{0}^{\infty} e^{-\epsilon/k_{\rm B}T} N(\epsilon)N(\epsilon+\hbar\omega) \, d\epsilon.
\end{equation}
The density of state $N(\epsilon)$ is defined by the retarded Green function $N(\epsilon)=\langle{\rm Re}G^{R}\rangle_{\rm FS}$, where $\langle...\rangle_{\rm FS}$ represents averaging over the Fermi surface.
The equations Eq.~(\ref{eq:Rs}) and (\ref{eq:sigma1}) are generally valid to estimate the loss caused by quasi-particles at a given temperature when the model Green function is given.

\section{Green function model of various superconducting materials}
In this section, we review various models of superconductors that give $G^{R}$.
We do not go through the exhaust list of all the previous theories 
but rather list the simplest possible toy-models that may provide didactic lessons on the general optical response.
For simplicity, we consider homogeneous superconductors in the quasi-classical approximation.

\subsection{Niobium and Nb$_{3}$Sn}
The quasi-classical Green function of the conventional s-wave superconductors is given by the BCS theory\footnote{There are a lot of textbooks and papers that introduce quasi-classical Green functions for instance Eq.(5.46) of Ref.~\cite{kopnin01}}
\begin{equation}\label{eq:Green_swave}
    G^R = \frac{\epsilon+i\delta}{\sqrt{(\epsilon+i\delta)^2-\Delta_0^2}}.
\end{equation}
The well-known formula Eq.~(\ref{eq:BCS}) is reproduced by substituting Eq.~(\ref{eq:Green_swave}) into Eq.~(\ref{eq:Rs}) and Eq.~(\ref{eq:sigma1}).
As mentioned above, the twice larger superconducting gap $\Delta_0$ differentiates Nb$_{3}$Sn from niobium.
$\delta$ is a parameter to avoid divergence during the numerical integral in Eq.~(\ref{eq:sigma1}).
This $\delta$ was considered in the Dynes model~\cite{PhysRevLett.41.1509}, in which $\Gamma$ is often used instead of $\delta$, and has been used to reproduce $R_{\rm res}$ phenomenologically~\cite{PhysRevApplied.17.014018} or by introducing the pair-breaking term in the effective Hamiltonian~\cite{PhysRevB.104.094519}.

\subsection{MgB$_{2}$}
MgB$_{2}$, discovered in 2001~\cite{2001Natur.410...63N},  has two superconducting gaps in $\sigma$-band and $\pi$-band.
Even the smaller $\pi$-gap is larger than the gap of niobium.
Due to its potential in higher accelerating gradients,
some groups have been experimentally investigating MgB$_{2}$ for SRF cavities~\cite{Xi_2009, PhysRevSTAB.17.012001, Tan_2016}.
The surface resistance of MgB$_2$ was measured in strip-line resonators~\cite{1440379} at 10~GHz.
The complex optical conductivity was measured in a Mach-Zehnder interferometer with submillimeter waves~\cite{PhysRevLett.87.097003} and calculated by for example Ref.~\cite{SHAHZAMANIAN2005103}.

We follow the simple model developed by Watanabe and Kita~\cite{doi:10.1143/JPSJ.73.2239}.
They considered two isotropic bands ($\sigma$- and $\pi$-bands) of MgB$_2$ and calculated inter-bands scattering.
They compared specific heat data with their theory but did not calculate optical conductivity.
Although this model we adopted in this work does not include nodes,
some experimental studies indicated nodes in $\pi$-gap of MgB$_2$, via direct measurement of $\sigma_1$~\cite{PhysRevLett.87.097003} and passive intermodulation~\cite{PhysRevB.80.174522}.
B.~Xiao and C.~E.~Reece reported calculation of surface resistance of MgB$_2$~\cite{PhysRevAccelBeams.22.053101} without the inter-band scattering.

The density of states is given by
\begin{equation}
   N(\epsilon) = \sum_{\alpha=\sigma, \pi} N_{\alpha}(0) {\rm Re} G^R_\alpha(\epsilon).
\end{equation}
If the inter-band scattering is disregarded, 
the Green functions are simply given by
\begin{equation}
    G^R_\alpha = \frac{\epsilon+i\delta}{\sqrt{(\epsilon+i\delta)^2-\Delta_\alpha^2}},
\end{equation}
with $\Delta_{\alpha}=\Delta_{\sigma}, \Delta_{\pi}$, the superconducting gaps of $\sigma-$ and $\pi-$bands, respectively.
Since we disregard the inter-band scattering for simplicity, 
the results may be comparable to the dirty limit case of Ref.~\cite{PhysRevAccelBeams.22.053101}.

\subsection{Cuprates}
The cuprates are the typical high-T$_{\rm c}$ superconductors discovered in 1986~\cite{1986ZPhyB..64..189B}.
The most characteristic property of cuprates is nodes in their gap as a d-wave superconductor.
The toy-model Green function is given by~\cite{coleman_2015}
\begin{align}
    G^{R} & =  \frac{\epsilon + i\delta}{\sqrt{(\epsilon + i \delta)^2 - \Delta^2(\theta)}} \\
    \Delta(\theta) & =  \Delta_0 \cos(2\theta)
\end{align}
Because of this gapless feature $\Delta_0 \cos(2\theta)$, 
cuprates suffer from more thermally excited quasi-particles, which are the cause of the RF loss considered in this paper.
Thus, despite their promising potential in the application for superconducting magnets, 
cuprates have not been seriously considered for the SRF cavities.
Recent studies on cuprate cavities are for dark matter axion searches~\cite{Golm:2021ooj, CAPP:2020utb, Ahn:2021fgb}.
In such an application, RF cavities are exposed to a strong static magnetic field with a very weak RF field.
The cuprate cavities offer one order of magnitude as a high-quality factor as copper cavities in such an environment.
The feasibility of the particle accelerator application is not fully understood.

\subsection{Iron-based pnictides}
The iron-based superconductors were discovered in 2008~\cite{doi:10.1021/ja063355c} and most of them show gap-full behavior.
Because of this gap-full feature, iron-based superconductors may be considered to be more promising than cuprates for the accelerating cavity application.
The gap structure does not contain nodes but is complex on the disconnected and anisotropic Fermi surface maybe due to the interplay among multiple bands~\cite{PhysRevLett.101.087004}.
In this paper, we follow the model developed by Nagai~\cite{Nagai_2008}.
He introduced the effective five-band model
and phenomenologically assumed anisotropic $\pm$s-wave pair function
\begin{align}
    &\Delta_{\alpha_{1,2}, \beta_{1,2}}  =  \Delta_0 \Phi_{\alpha_{1,2}, \beta_{1,2}} \\
    &\Phi_{\alpha_{1,2}} = \Phi_{\rm a} \\
    &\Phi_{\beta_{1,2}} = \frac{1+\Phi_{\beta_{\rm min}}}{2} \pm \frac{1-\Phi_{\beta_{\rm min}}}{2} \cos{\left(2\phi_{1,2} \right)}.
\end{align}
Then, the Green function is given by
\begin{align}
    G^{R} & =  \frac{\epsilon + i\delta}{\sqrt{(\epsilon + i \delta)^2 - \Delta^2_{\alpha_{1, 2}, \beta_{1,2}}(\phi_{1,2})}}
\end{align}
with the parameters of the model ($\Delta_0$, $\Phi_{\rm a}$, $\Phi_{\beta_{\rm min}}$)
that define the two-gap structure.
By using this model~\cite{Nagai_2008},
nuclear spin-lattice rate and superfluid density were evaluated.
However, optical conductivity was not calculated.

\subsection{Summary of density of states}
\begin{figure}[!t]
\centering
\includegraphics[width=4in]{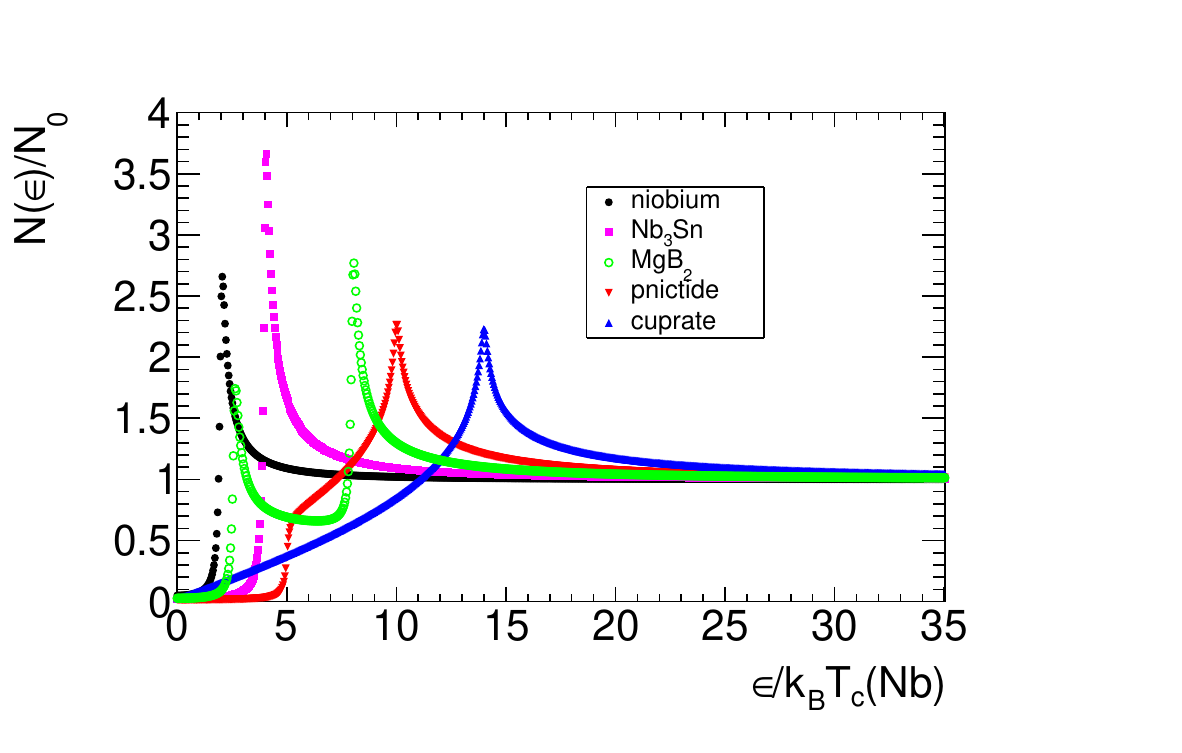}
\caption{Comparison of density of states of the materials modeled by Green functions phenomenologically.}
\label{fig:DOS}
\end{figure}

\begin{table*}[!t]
\renewcommand{\arraystretch}{1.3}
\caption{Summary of material parameters}
\label{tab:param}
\centering
\begin{tabular}{cccccccc}
\hline
material & $T_{\rm c}$ [K] & $\Delta_0/k_{\rm B}T_{\rm c}$ & $\delta/k_{\rm B}T_{\rm c}$ & $\lambda$ [nm] & $\Phi_{\rm a}$ & $\Phi_{\beta_{\rm min}}$ & $N_{\sigma}/N_{\pi}$\\
\hline
niobium & 9.25 & 2 & 0.1 & 40 & - & - & - \\
Nb$_3$Sn & 18.3 & 2 & 0.1 & 80 & - & - & - \\
MgB$_2$ & 39 & 2/0.65 & 0.1 & 140 & - & - & 0.72 \\
pnictide & 50 & 2 & 0.1 & 200 & 1 & 0.5 & - \\
cuprate & 70 & 2 & 0.1 & 150 & - & - & - \\
\hline
\end{tabular}
\end{table*}
By taking the average over the Fermi surface and extracting the real part, 
we obtain the density of states of each material model from the Green functions introduced above.
For simplicity, we take the quasi-classical approximation, assuming a spherical or cylindrical Fermi surface without complex band-structure calculation.
Figure~\ref{fig:DOS} shows the example of the density of states that we consider in this paper.
The parameters assumed in this example are listed in Table~\ref{tab:param}.

The density of states of cuprate shows a qualitatively different behavior at low energy due to its d-wave nature.
This gives rise to more quasi-particles in the gap than s-wave superconductors and thus may have more RF loss.
We need to perform numerical integral to quantify this argument
because the literature data has been often dominated by vortex loss under strong magnetic fields~\cite{Ahn:2021fgb}, 
not what we are aiming at in this study.
All the other s-wave density of states exhibit sub-gap states below the smallest gap, 
which are introduced by the parameter $\delta$ for numerical calculation, smearing the pole at the gaps as well.
One can argue that this $\delta$ may vary depending on magnetic impurities.
Here in this study, in order to reveal the most fundamental aspects of the RF loss,
we keep $\delta$ constant for all the models.

\section{Results}
\subsection{Optical conductivity}
\begin{figure}[!t]
\centering
\includegraphics[width=4in]{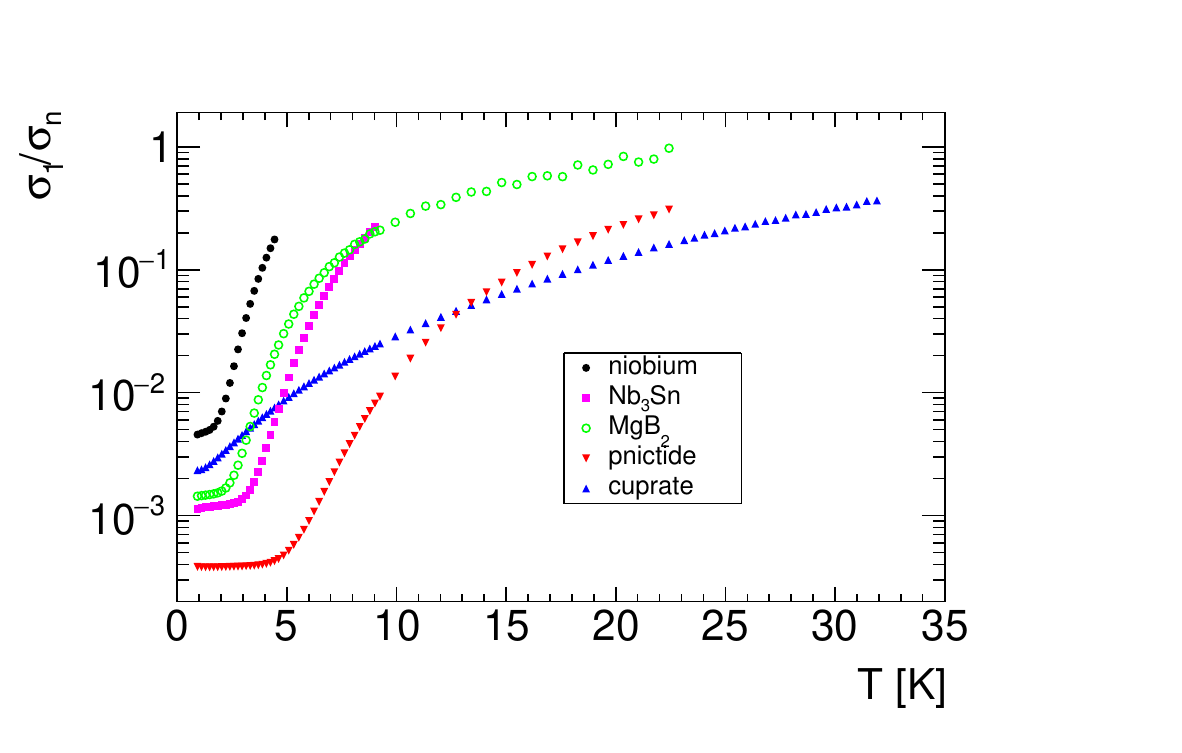}
\caption{Comparison of the real part of the optical conductivity $\sigma_1$ (lossy component).}
\label{fig:sigma_1}
\end{figure}

By performing the integral in Eq.~(\ref{eq:sigma1}) numerically, we obtain the real part of the optical conductivity as shown in Fig.~\ref{fig:sigma_1}.
We fix the angular frequency $\hbar\omega/k_{\rm B}T_{\rm c}{\rm (Nb)}$ at 0.02, 
corresponding to around 600~MHz, which is sufficiently lower than any gaps in the models
\footnote{This frequency range is of particular interest for proton linear accelerators and energy frontier circular colliders}.
The improvement of $\sigma_1/\sigma_{\rm n}$ by cooling down is saturated at low temperature due to the parameter $\delta=0.1$ for the numerical integral.

Except for the cuprate, by using three parameters $(A, \Delta, B)$ the plots can be well fitted by an analytical formula for one-gap s-wave superconductors~\cite{halbritter70}
\begin{equation}
    \frac{\sigma_1}{\sigma_{\rm n}} = \frac{A}{T} \exp{\left(-\frac{\Delta}{T} \right)} + B,
\end{equation}
even though MgB$_2$ and pnictide have the two-gap structure.
Naively speaking, the smaller gap dominates the optical conductivity 
thanks to the exponential behavior of the Fermi Dirac distribution function $\exp{(-\epsilon/k_{\rm B}T)}$.

The cuprate results can be fitted by
\begin{equation}
    \frac{\sigma_1}{\sigma_{\rm n}} = C T^{\alpha} + B,
\end{equation}
with the best fitting power of $\alpha=2.34\pm0.01$.
This power law around $\alpha=2$ was obtained by more complete calculations~\cite{PhysRevLett.71.3705, PhysRevB.50.10250}.
In this work, we could reproduce their results in a much simpler numerical integral of the toy-model Green function.

With the parameters we consider in Table~\ref{tab:param}, 
pnictide shows the lowest ratio $\sigma_1/\sigma_{\rm n}$ below 10~K.
This is an encouraging result for further development of the theory and technology.
However, one needs to be careful about the penetration depth as we discuss in the next section.

\subsection{Surface resistance of bulk materials}
\begin{figure}[!t]
\centering
\includegraphics[width=4in]{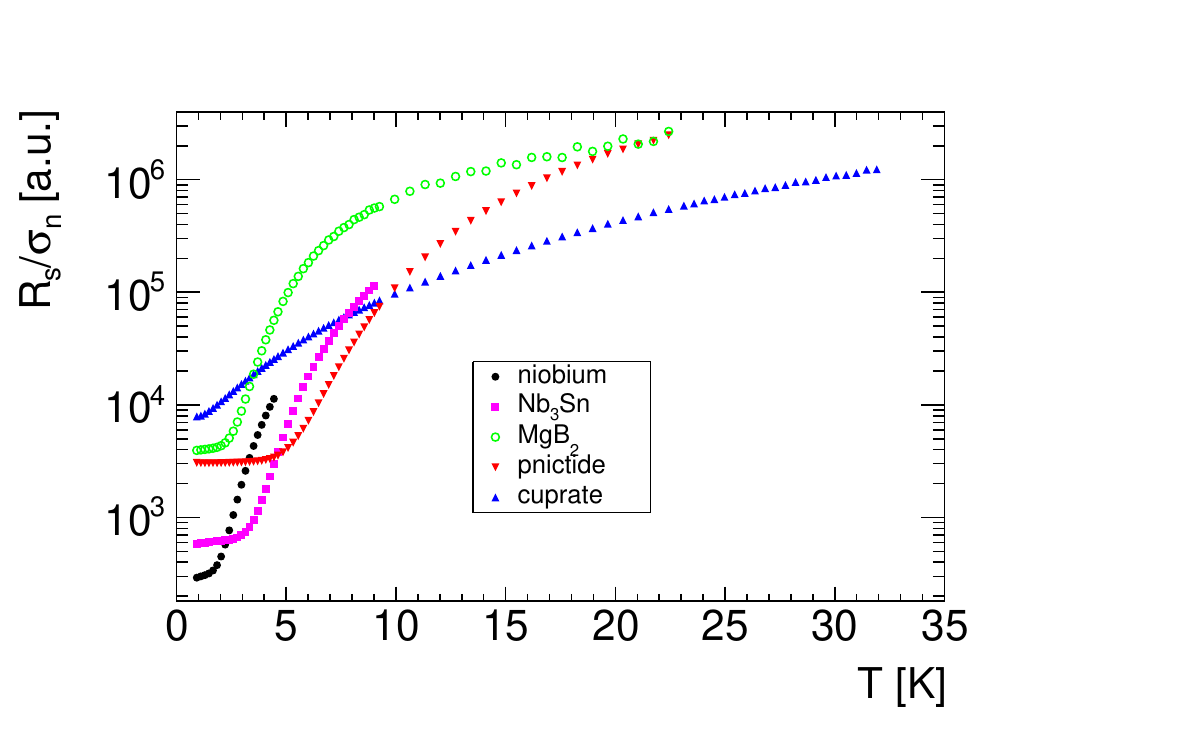}
\caption{Comparison of $R_{\rm s}$ of bulk materials normalized by normal conducting DC conductivity $\sigma_{\rm n}$.}
\label{fig:Rs}
\end{figure}

In order to predict surface resistance in the SRF cavities,
the $\sigma_1$ results need to be integrated over the RF field penetrating into the material.
When we consider a bulk material in the dirty limit, 
the surface resistance can be simply approximated by Eq.~(\ref{eq:Rs}).
In Fig.~\ref{fig:Rs}, we compare $R_{\rm s}/\sigma_{\rm n}$ by multiplying $\lambda^3$ listed in Table~\ref{tab:param}\footnote{
The listed values are for London penetration depth and strictly speaking are not immediately relevant in the dirty limit case that this paper is focusing. In the dirty limit, the penetration depth is extended by more than factor 1.6 but is common to all the materials; thus, the ratios discussed in this paper are not strongly influenced.
} to $\sigma_1/\sigma_{\rm n}$ in an arbitrary unit.
Note that $\sigma_{\rm n}$ is still kept as a constant of each material.

We have several observations from these results.
First, although niobium showed the highest $\sigma_1/\sigma_{\rm n}$ in Fig.~\ref{fig:sigma_1}, 
its short $\lambda$ leads to the lowest $R_{\rm s}$ at temperature lower than 3~K.
Secondly, the surface resistance of Nb$_3$Sn at 4~K is comparable to that at niobium at 2~K,
which is the same conclusion by using the analytical formula Eq.~(\ref{eq:BCS}).
Thirdly, although pnictide has the lowest $\sigma_1/\sigma_{\rm n}$,
its long $\lambda$ leads to relatively large $R_{\rm s}/\sigma_{\rm n}$ below 5~K.
Finally, although cuprate shows largest $R_{\rm s}/\sigma_{\rm n}$ at the low temperature due to its gapless nature and corresponding quite a few number of quasi-particles,
it gives the best $R_{\rm s}/\sigma_{\rm n}$ over 10~K thanks to the high T$_{\rm c}$.

\section{Discussion}
\subsection{Thin film application}
This study revealed that the surface resistance of bulk superconductors beyond Nb$_3$Sn may need further consideration due to their long penetration depth.
Developing a film thinner than $\lambda$ and/or a multi-layer structure may be a natural extension.
For technical realization of fabricating cavities with materials beyond niobium, 
cavities will be anyway based on thin-film technology~\cite{antoine:ipac2023-moyd3}.

The niobium thin-film technology was pioneered in the Large Electron Positron (LEP) accelerator,
and then adopted in the Large Hadron Collider (LHC), SOLEIL, Acceleratore Lineare Per Ioni (ALPII), and recently in HIE-ISOLDE.
The most mature methodology is niobium sputtering on copper substrates.
The researches on unexpected nonlinear surface resistance (Q-slope problem) have been widely investigated for the niobium film cavities~\cite{benvenuti99, calatroni16, PhysRevAccelBeams.22.073101}; 
however, a complete understanding and solution of this issue have not yet been obtained.

There have been rich R\&D on Nb$_3$Sn thin film for further improvement beyond niobium.
Although Nb$_3$Sn sputtering on copper substrates is being investigated~\cite{Ilyina_2019}, currently, the most successful technology is based on Sn-diffused alloy formation on niobium substrates~\cite{Posen_2017}.
The present limitation is maximum field reach around 22~MV/m in an elliptical cavity at 1.3~GHz and the cause has not been fully understood.

Both niobium and Nb$_3$Sn films are, however, thicker than their penetration depths; 
thus, they may in principle react like bulky material against microwaves.
The practical issues listed above may not be a direct consequence of the film thickness.

More recent R\&D focuses on multi-layered structure~\cite{10.1063/1.2162264, Kubo_2017}.
The key concept is that the layer thick is thinner than the penetration depth.
Being separated by dielectric layers,
penetrating RF vortices would be unstable so that the gradient reach may be improved.
The experimental evidence on a higher critical field is reported via third harmonics measurement~\cite{katayama:srf2019-thfua2} and muon spin rotation~\cite{PhysRevAccelBeams.21.032002}.

The possible application of the non-conventional superconducting materials may be in line with this multi-layer structure~\cite{Antoine_2019}.
A potential of higher gradient by ion-based superconductors was briefly mentioned in Ref.~\cite{Gurevich17}.
Recently, an encouraging result on a higher critical field was obtained on FeSe films on niobium substrate~\cite{Lin_2021}.
An open question is the RF loss of such a thin film made of non-conventional superconductors.
This study showed that the iron-based pnictide may be an ideal candidate to be at the top layer thanks to its potentially low optical conductivity $\sigma_1$.
Combined analysis of the superheating field and RF loss will be future work.

\subsection{Medium pulse operation}
Bulk niobium cavities are operated in the long pulse (ms) to Continuous Wave (CW) accelerators at 2~K or 4~K.
Normal conducting copper cavities are operated in the short pulse (${\rm \mu}$s) accelerators usually at room temperature.
Since the materials investigated in this paper have intermediate performance between superconducting niobium and copper, 
one could think of medium pulsed (10-100~${\rm \mu}$s) accelerators at temperature higher than 10~K.

For example, according to our results in Fig.~\ref{fig:Rs},
cuprate above 15~K may have surface resistance only factor 100 higher than that of niobium below 4~K.
Since the state-of-the-art niobium cavities have the intrinsic quality factor of $Q_0 \sim10^{10}$,
let us assume that future cuprate cavities may be able to reach $Q_0\sim10^8$.
This may indicate a potential operation with a pulse length of 10~${\rm \mu}$s, shortened from 1~ms of niobium cavities by two orders of magnitude.
Even though superconducting cavities themselves have a high $Q_0$ with slow response ($>$ms),
for high peak current machines, the response time of the cavities $\tau_{\rm L}$ is dominated by external quality factor $Q_{\rm ext}$ of high-power couplers whose energy is coupled to the beam.
This can be seen by introducing the loaded quality factor $Q_{\rm L}$ as follows
\begin{align}
    & \tau_{\rm L} = \frac{Q_{\rm L}}{\omega} \\
    & \frac{1}{Q_{\rm L}} = \frac{1}{Q_{0}} + \frac{1}{Q_{\rm ext}}.
\end{align}
The $Q_{\rm ext}$ has optimal value depending on the beam current $I_{\rm b}$ and cavity voltage $V_{\rm c}$
\begin{equation}
    Q_{\rm ext} = \frac{V_{\rm c}}{(R/Q)I_{\rm b} \cos{\phi_{\rm s}}},
\end{equation}
with $R/Q$ a geometrical constant representing shunt impedance over quality factor, and $\phi_{\rm s}$ the synchronous phase.
For $V_{\rm c}=5\, {\rm MV}$, $I_{\rm}=200\, {\rm mA}$, $R/Q=300, {\rm \Omega}$, and $\phi_{\rm s}=20^\circ$,
\begin{align}
    & Q_{\rm ext} \sim 8.3\times10^4 \ll Q_0\\
    &\tau_{\rm L}\sim \frac{Q_{\rm ext}}{\omega} \sim 13\, {\rm \mu s}.
\end{align}
Thus, in principle, a medium pulse beam may be an option for such a new class of cavities made of high-T$_{\rm c}$ superconductors.

\subsection{Other sources of loss} \label{sec:residual}
The cause of surface resistance considered in this study is thermally excited quasi-particles.
This is the main source of RF loss in the Meissner state of clean materials at finite temperatures.
However, there are many other sources of loss, for instance
\begin{itemize}
    \item nonlinear Meissner effect due to anisotropic pairing
    \item sub-gap states due to impurities
    \item non-equilibrium distribution functions
    \item oscillation of trapped flux
    \item Josephson junction (weak link model)
    \item normal conducting defect
    \item heating effect
\end{itemize}
For the axion dark matter experiment,
cavities are operated in the mixed state under a strong magnetic field; 
thereby the loss is dominated by the oscillation of the trapped flux.
The relevance of these additional losses in the accelerating cavities is still an open question.
For example, one can argue that potentially multi-metallic structure of the new materials inevitably introduce thermoelectric currents to be trapped, in a way similar to or even more complex than the observations in Nb-coated copper~\cite{PhysRevAccelBeams.22.073101} and Nb$_3$Sn-alloy formed niobium cavities~\cite{Hall:2017vvs}.

Another well-known example of additional RF loss was observed in cuprates in the 1990s. 
This was explained by the weak link~\cite{PhysRevB.38.6538, PhysRevB.43.6128} unlike that of niobium.
The small grains and short coherence length were well modelled by Josephson junctions and gave rise to the substantial residual resistance.
It is of great importance to revise this old results in the most advanced cuprate samples with the recent technology.
A similar analysis on recent ion-based superconductors would also be motivated.

Another potential research direction is an extension of Dynes model in non-conventional superconductors.
In this study, an additional loss caused by the sub-gap states was unintentionally included through the parameter $\delta$ to avoid poles in the numerical integrals.
Quantitative physics interpretation of $\delta$ over different materials is clearly beyond the scope of this paper; however, we can still consider possible studies for the future.
In s-wave superconductors, phenomenological approaches, supported by the relevant theory background~\cite{PhysRevB.94.144508} imply that the Dynes parameter is related to magnetic (pair breaking) impurities in the material.
One could imagine that the ion-based superconductors might suffer from higher Dynes parameter due to the intrinsically magnetic property of the ion elements in the crystal structure.
The most astonishing property of ion-based superconductors is that such potentially magnetic material becomes also superconducting.
It would be important to study remaining magnetic effect and its impact to the microwave response if the simple Dynes model is still valid. 

\section{Conclusion}
We generalized the well-known analytical formula of surface resistance to be compatible with non-conventional superconductors.
Based on several toy-models of Green functions in the literature, 
we demonstrated a simple numerical calculation of optical conductivity and bulk surface resistance.
Even though our approach is oversimplified, some previously known features were reproduced.
The results indicated some opportunities in thin-film cavities as well as medium pulse accelerators.
It is of importance to investigate the state-of-the-art materials, starting from the most idealized models of loss mechanism presented in this paper.
This paper provided a systematic mean to compare different superconducting materials and implied future research directions in the SRF application.


%



\section*{Acknowledgment}
This work was inspired by discussions with Sergio Calatroni and Claire Antoine.
The author would like to thank Hikaru Ueki, Yuki Nagai, Takafumi Kita, Daniel Oates, and Takayuki Kubo for the valuable discussions. Special thanks goes to Nicola Pompeo who invited me to the 15th International Workshop On High Temperature Superconductors in High Frequency Fields, in which the discussions were initiated.

\ifCLASSOPTIONcaptionsoff
  \newpage
\fi



\bibliographystyle{IEEEtran}
\bibliography{IEEEabrv,SRF_HTS}
\end{document}